\documentclass{pasj00}

\begin{document}
\SetRunningHead{Author(s) in page-head}{Running Head}

\title{All-sky video orbits of Lyrids 2009}

\author{Juraj \textsc{T\'{o}th}, Leonard \textsc{Korno\v{s}}, Peter {\sc Vere\v{s}}, Ji\v{r}\'{i} {\sc \v{S}ilha}}
\author{Du\v{s}an {\sc Kalman\v{c}ok}, Pavol {\sc Zigo} and Jozef {\sc Vil\'{a}gi}}
\affil{Faculty of Mathematics, Physics and Informatics, Comenius
University in Bratislava, \\ Mlynsk\'{a} dolina, 842 48
Bratislava, Slovak Republic} \email{toth@fmph.uniba.sk}

%


\maketitle

\begin{abstract}
We report observational results of the Lyrid meteor shower
observed by the double station all-sky video system in the night
of April 21/22, 2009 at the Astronomical and Geophysical
Observatory of the Comenius University in Modra and Arboretum,
Tes\'{a}rske Mly\v{n}any, Slovakia. This observation was the first
test of the double stations and orbit determination method within
the frame of the new Slovak Video Meteor Network (SVMN). We
present the whole set of 17 observed orbits of Lyrids as well as
the five most precise orbits in detail form. The comparison with
the known datasets, precise photographic IAU MDC and SonotaCo
video orbits, demonstrate quite good consistency and similar
quality.
\end{abstract}

\section{Introduction}

The fish-eye video meteor system at Astronomical and Geophysical
Observatory (AGO) in Modra, Slovakia, has started regular
observations on April 1, 2007. The system was originally developed
at our institute and consists of a fish-eye Canon 2.4/15\,mm
objective, 2'' Mullard image intensifier, Meopta 1.9/16\,mm lens.
The observation presented here were done by the Watec 120N camera.
The analog video signal is digitized in the real time and analyzed
by the UFOCapture software (author SonotaCo, {\it
http://sonotaco.com/e\_index.html}), which is able to detect any
moving objects including meteors. The resolution of the system is
$720\times540$ pixels (15 arcmin/px), corresponding to a field of
view of $170{^\circ}\times 140{^\circ}$. The limiting stellar
magnitude is $+5.5^{m}$ and meteors up to the magnitude $+3.5^{m}$
are detected. The astrometric precision of reference stars is in
average $\sim$ 5 arcmin by using few hundreds reference stars. The
system operates autonomously (\cite{toth}).

The second station, at the time of the observation equipped with
the same opto-electronical system and a Watec 902 H2 analog
camera, is located at the Arboretum of the Slovak Academy of
Sciences, Tes\'{a}rske Mly\v{n}any, 80 km in the East direction
from Modra. The second station is semiautomatic and is controlled
through the internet (remote network access). Both video stations,
Modra and Arboretum, constitute the base of the new developing
Slovak Video Meteor Network.

\section{Observations and data analysis}

The observation of Lyrid meteor shower, at the night of April
21/22, 2009 from 19:15 to 2:20 UT, was the first observational
test of double stations operation and following orbit calculation.
We obtained reliable observational data covering a substantial
part of the maximum of Lyrid's activity. The data were analyzed by
the UFO Analyzer and the UFO Orbit software ({\it
http://sonotaco.com/e\_index.html}). We detected 78 and 52 meteors
from the first and the second station, respectively. 32 meteors
were simultaneously observed at both stations, 17 of them were
identified as Lyrids.

\begin{figure*}
  \begin{center}
    \FigureFile(80mm,60mm){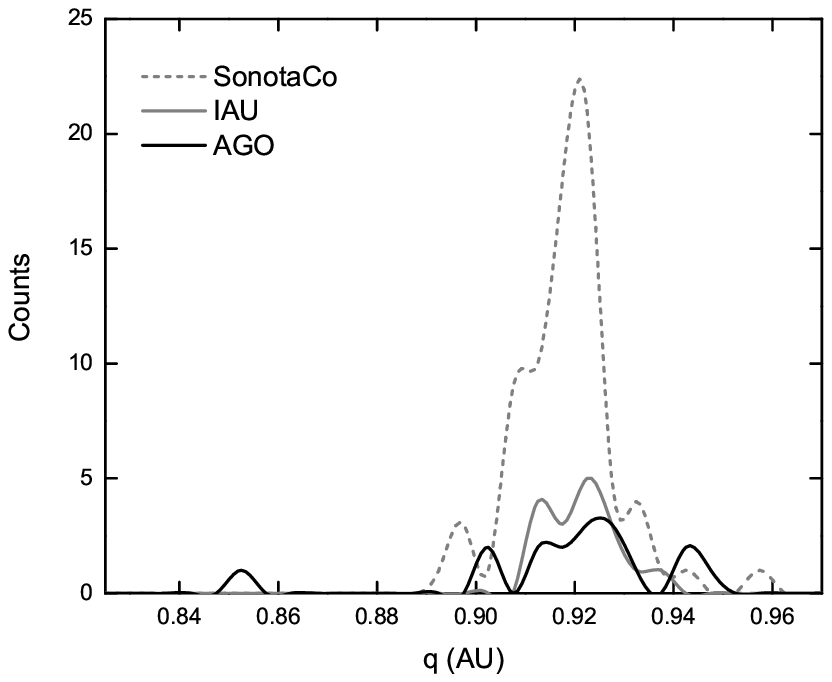}
    \hspace{0.3cm}
      \FigureFile(80mm,60mm){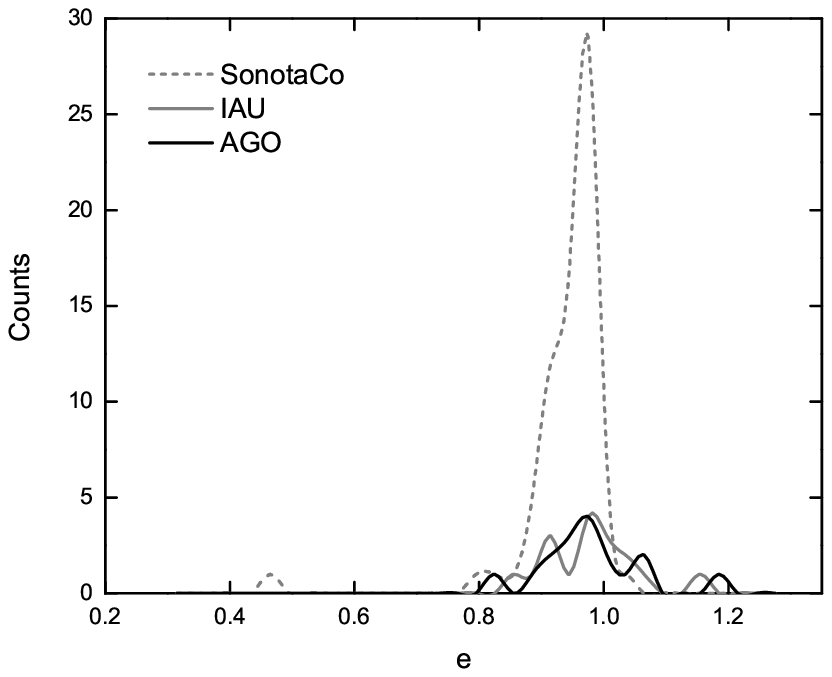}
    \vspace{0.7cm}
      \FigureFile(80mm,60mm){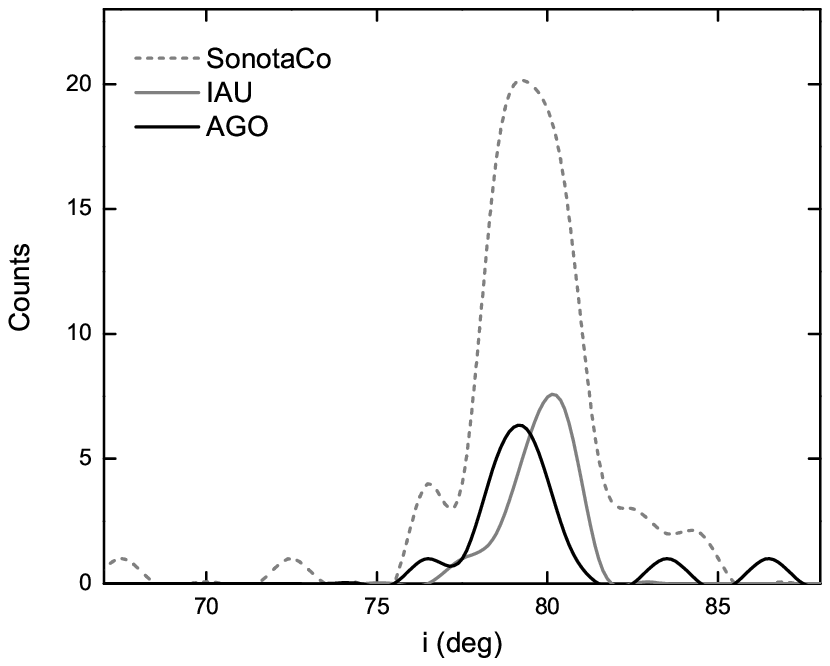}
    \hspace{0.3cm}
      \FigureFile(80mm,60mm){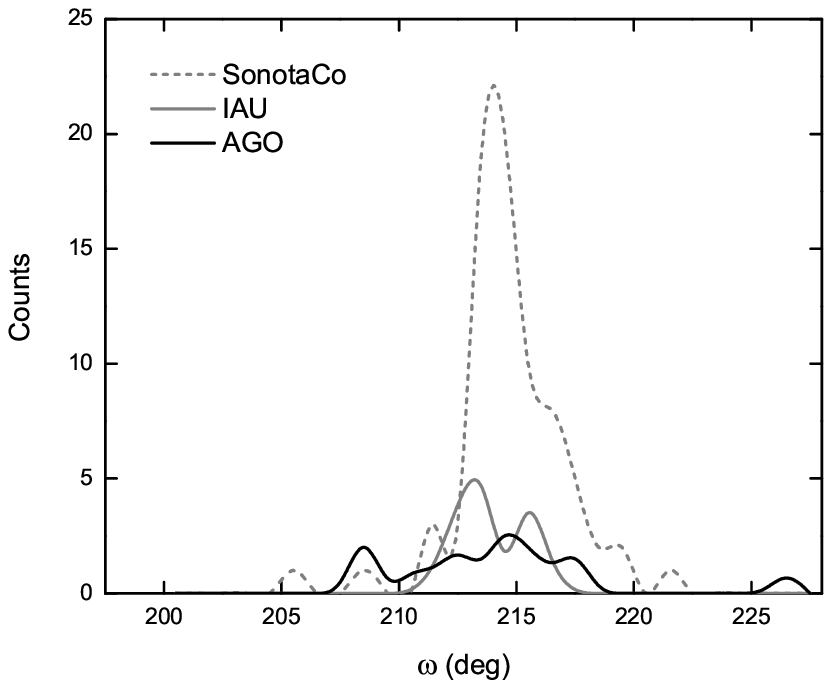}
 \vspace{0.7cm}
      \FigureFile(80mm,60mm){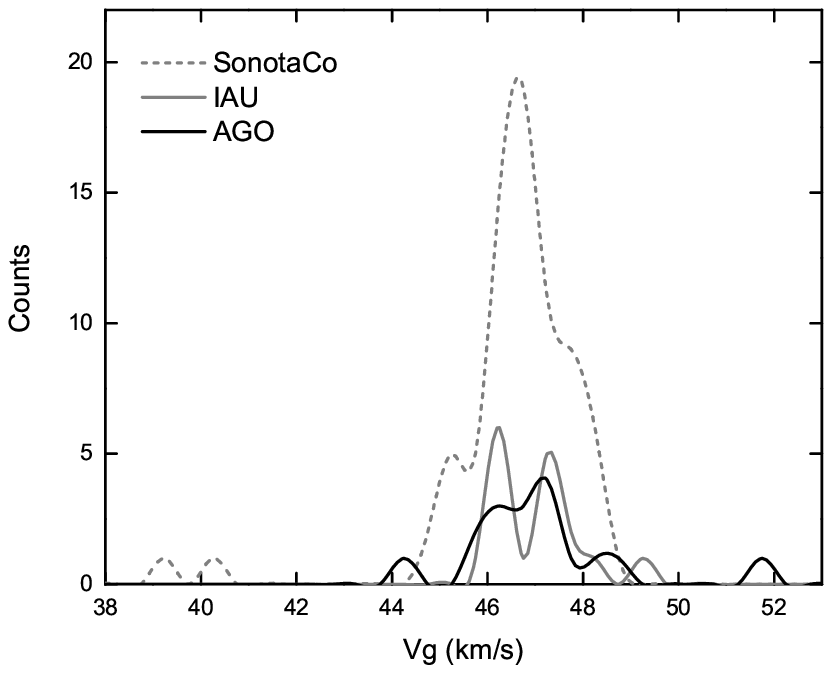}
    \hspace{0.3cm}
      \FigureFile(80mm,60mm){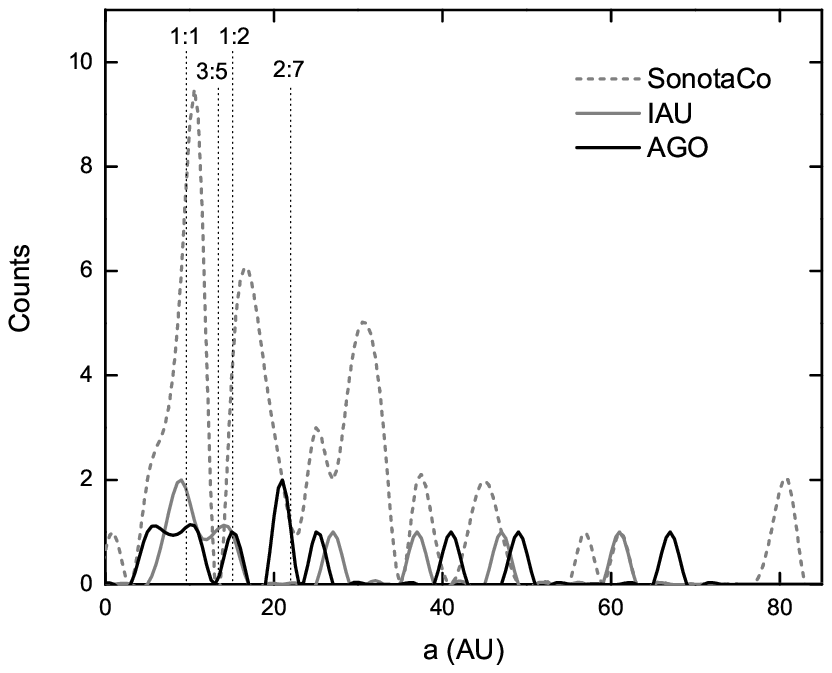}
  \end{center}
  \caption{The distribution of the orbital elements (eq. 2000.0) of Lyrid meteors (q, e, i, $\omega$, a) as well
  as the geocentric velocity of precise IAU MDC photographic orbits (\cite{lind}),
  SonotaCo Japanese database (http://sonotaco.jp/doc/SNM/) for 2007-2009 and
  observations from Modra -- Arboretum (AGO), 2009. The semimajor axis graph contains mean motion resonances with Saturn.}
\label{fig:elm}
\end{figure*}

\begin{longtable}{lrrrrrrrrrr}
  \caption{Mean values and standard deviations of the orbital elements, geocentric radiants (eq. 2000.0) and velocities for the short ($<$200 years),
  long ($>$200 years) periodic and hyperbolic subset of observed Lyrid meteors on the base of Modra -– Aboretum, April 21/22, 2009.
  The same parameters of the parent comet C/1861 G1 (Thatcher) are displayed for the comparison (\cite{mar}).}
  \label{tab:tab1}
  \hline
   & a (AU) & q (AU) & e & i ($^\circ$) & $\omega$ ($^\circ$) & $\Omega$ ($^\circ$) & $\alpha$ ($^\circ$) & $\delta$ ($^\circ$) & $V_{g}$ (km/s) & n \\
\endfirsthead
  \hline
\endhead
  \hline
\endfoot
  \hline
\endlastfoot
  \hline
short-periodic & 14.46      & 0.912      & 0.919      & 78.7     & 216.0    & 31.9     & 271.3    & 33.2     & 46.07     & 8  \\
               & $\pm$7.16  & $\pm$0.026 & $\pm$0.044 & $\pm$0.9 & $\pm$4.7 & $\pm$0.1 & $\pm$1.9 & $\pm$1.5 & $\pm$0.76 &    \\
long-periodic  & 52.90      & 0.919      & 0.982      & 79.0     & 214.1    & 31.9     & 271.3    & 33.7     & 46.77     & 3  \\
               & $\pm$13.50 & $\pm$0.006 & $\pm$0.004 & $\pm$1.1 & $\pm$1.1 & $\pm$0.1 & $\pm$1.0 & $\pm$0.3 & $\pm$0.48 &    \\
hyperbolic     & --         & 0.930      & 1.065      & 81.1     & 211.2    & 31.8     & 272.6    & 33.5     & 48.39     & 6  \\
               &            & $\pm$0.018 & $\pm$0.062 & $\pm$3.1 & $\pm$3.7 & $\pm$0.1 & $\pm$2.4 & $\pm$1.6 & $\pm$1.75 &    \\
comet Thatcher & 55.62      & 0.921      & 0.984      & 79.8     & 213.5    & 31.9     & 272.0    & 33.5     & 47.08     &    \\
 \end{longtable}

\begin{longtable}{rrrrrrrrrrr}
  \caption{Mean values of the orbital elements, geocentric radiants (eq. 2000.0) and velocities and the absolute magnitudes of the five most
  precise orbits of observed Lyrids on the base Modra -– Arboretum, April 21/22, 2009.}
  \label{tab:tab2}
  \hline
   Date-Time & a (AU) & q (AU) & e & i ($^\circ$) & $\omega$ ($^\circ$) & $\Omega$ ($^\circ$) & $\alpha$ ($^\circ$) & $\delta$ ($^\circ$) & $V_{g}$ (km/s) & $M_{abs}$ \\
\endfirsthead
  \hline
\endhead
  \hline
\endfoot
  \hline
\endlastfoot
  \hline
 20090421 221532 & 7.44   & 0.922  & 0.876 & 79.4 & 214.6  & 31.80488  & 272.8 & 33.1 & 45.96 & +0.0  \\
 20090422 005955 & 9.98   & 0.922  & 0.908 & 78.5 & 214.3  & 31.91625  & 272.1 & 33.7 & 45.84 & --1.5  \\
 20090422 010531 & 49.83  & 0.914  & 0.982 & 77.7 & 215.2  & 31.92006  & 270.2 & 34.1 & 46.21 & --2.9  \\
 20090422 013515 & 67.66  & 0.918  & 0.986 & 79.5 & 214.3  & 31.94020  & 271.5 & 33.4 & 47.02 & --2.5  \\
 20090422 015213 & 20.40  & 0.930  & 0.954 & 78.0 & 212.2  & 31.95169  & 272.1 & 34.6 & 46.05 & --3.1  \\
  \end{longtable}

The orbits of Lyrids from Modra -- Arboretum are consistent with
those previously derived by several authors (see \cite{jen}, p.
702) and are presented in figure \ref{fig:elm} (black line). The
orbital element distributions are depicted by using a B-spline
technique. The observed 17 Lyrids are compared with 17 IAU MDC
photographic orbits (\cite{lind}) and 75 Lyrids from the SonotaCo
Japanese database of video orbits (http://sonotaco.jp/doc/SNM/)
obtained in 2007 - 2009. The orbits from SonotaCo database
represent the most precise subset of Lyrids in the database
selected by the high quality criteria (\cite{ver}).

Nevertheless, there are  some hyperbolic orbits in all three
datasets. The IAU Meteor Database contains 35\%, SonotaCo 8\% and
our data 35\% Lyrids on hyperbolic orbits. However, the
hyperbolicity of meteors might not be real. According to
\citet{haj}, the most probable reason is the uncertainty in
velocity determination, shifting a part of the data through the
parabolic limit. Therefore, only elliptic orbits in the
distribution of semimajor axis in figure \ref{fig:elm} are
presented.

Analogically to \citet{por}, we have divided the observed Lyrids
into three groups, short periodic ($<$ 200 years), long preriodic
($>$ 200 years) and hyperbolic orbits. The mean orbital elements
are presented in table \ref{tab:tab1}. Their standard deviations
are rather small except for semimajor axis, which is very
sensitive for a precise determination of the geocentric velocity.
The long periodic part of the stream is very close to the orbit of
the parent comet Thatcher within the standard deviation intervals.
Our data are consistent and of similar quality as other data sets
of Lyrids (figure \ref{fig:elm}, or \cite{kot}). Table
\ref{tab:tab2} presents five most precise Lyrid orbits in detail.
The high quality criteria were adapted according to \citet{ver}.

\subsection{Radiant positions}

\begin{figure}
  \begin{center}
    \FigureFile(80mm,60mm){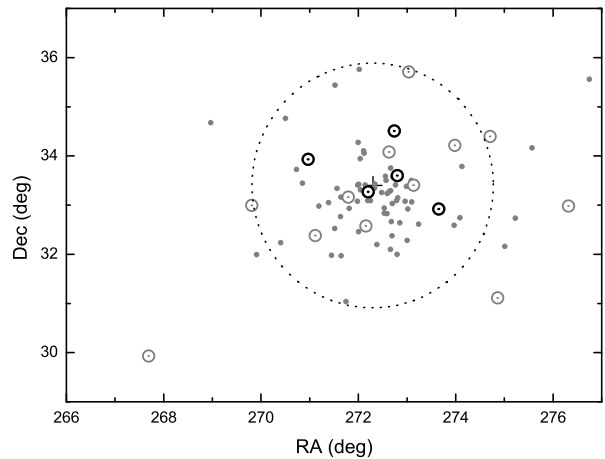}
  \end{center}
  \caption{Radiant distribution in the equatorial system (eq. 2000.0) of individual 17 Lyrid meteors ($\odot$) observed from Modra
-- Arboretum, April 21/22, 2009. Gray small dots are radiant
positions from SonotaCo data. Expected radiant position and 5
degrees circular area are depicted. The data are shown with
respect to the solar longitude 32.5$^\circ$.} \label{fig:rad}
\end{figure}

The radiant positions of individual Lyrids from Modra -- Arboretum
and from SonotaCo database are presented in figure \ref{fig:rad}.
The mean radiant derived from our 17 Lyrids for the mean solar
longitude of the time of the observation, $31.9^\circ$, is
following: $RA = 271.8^\circ \pm 2.0^\circ$, $Dec = 33.4^\circ \pm
1.3^\circ$.

For the comparison with the other authors, we recalculated
obtained mean radiant position to solar longitude $32.5^\circ$,
the usual time of the Lyrid's maximum activity, using equation
(1). Then, our radiant position is $272.6^\circ$, $33.2^\circ$,
which is consistent with the SonotaCo radiant $RA=272.4^\circ \pm
1.2^\circ$, $Dec=33.2^\circ \pm 0.8^\circ$ from 75 Lyrids as well
as with the radiant from IAU MDC (\cite{por}).

The observed number of 17 Lyrids was not sufficient for a daily
motion determination of the Lyrid meteor shower, that is why we
used 75 most precise Lyrids from SonotaCo database with individual
radiant information. The equation (1) describes this motion in
right ascension and declination:

\begin{equation}
\begin{array}{l}
\displaystyle RA=272.4^\circ + 1.25~(L_{\odot}-32.5^\circ)\\
\displaystyle DC=~33.2^\circ - 0.22~(L_{\odot}-32.5^\circ)
\end{array}
\label{eq:xdef}
\end{equation}

\noindent where 1.25$^\circ \pm 0.26^\circ$, and -0.22$^\circ \pm
0.18^\circ$ is the daily motion in RA and Dec, respectively. In
comparison with the work of \citet{por}, the motion in RA is a bit
higher from SonotaCo data.

\subsection{Beginning and terminal heights of Lyrids}

\begin{figure}
  \begin{center}
    \FigureFile(80mm,60mm){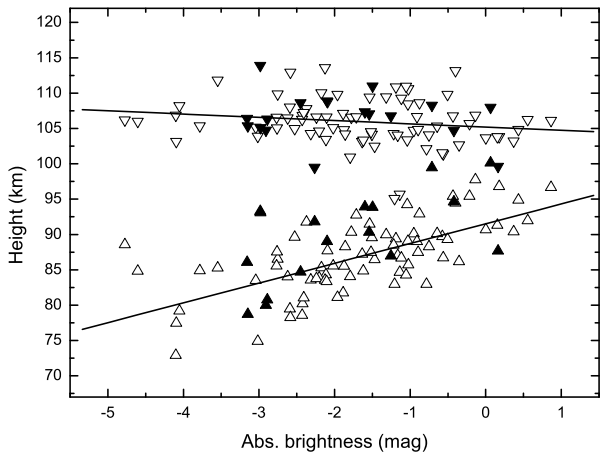}
  \end{center}
  \caption{The beginning and the terminal heights of 75 SonotaCo Lyrids (white triangles) and 17 Lyrids from Modra and Arboretum (black triangles)
  as a function of the absolute brightness.}
\label{fig:heig}
\end{figure}

The specific physical characteristics of the meteors, the
atmosphere interaction, the beginning and the terminal heights as
a function of the absolute brightness of Lyrids (figure
\ref{fig:heig}), are studied. This relation has not been inspected
yet. On the picture, there are depicted SonotaCo data and Modra
and Arboretum data matching quite well. When we have investigated
the heights of Lyrids in our data set, they behave similarly to
SonotaCo data set. For the improvement of our results from the
statistical reasons, we decided to show the beginning and terminal
heights as a function of the absolute brightness from both data
sets, which follows the equations:

\begin{equation}
\begin{array}{l}
\displaystyle H_{B}=105.2(\pm 0.6) - 0.5(\pm 0.3)~M_{A} \\
\displaystyle H_{E}=~~92.2(\pm 0.8) + 2.8(\pm 0.4)~M_{A},
\end{array}
\label{eq:hei}
\end{equation}

\noindent where $H_{B}$ stands for the beginning height, $H_{E}$
for the terminal height and $M_{A}$ for the absolute brightness.
Beginning heights almost do not depend on the absolute brightness,
however, terminal heights decrease with the increasing brightness.
\citet{kot1} showed, that beginning heights of Perseids, Leonids,
Orionids and Taurids increase as a function of the photometric
mass. On the contrary, beginning heights of Geminids almost do not
change with the photometric mass. According to \citet{kot1}, all
Geminids start to ablate in about 100 km, meaning, that their
meteoroids are more rigid and ablate near the melting point of the
silicates. It is accepted that Lyrids are of cometary origin but
their beginning heights behave in the similar way to Geminids.

\section{Discussion}

It seems that previous division of Lyrids for short and long
periodic orbits would need some revision. According to the
SonotaCo data, the semimajor axis distribution shows possible
resonant effects. The first and second peak (figure \ref{fig:elm},
lower right) are close to 1:1 (9.6 AU) and 1:2 (15.1 AU) mean
motion resonances with Saturn. Also the first two gaps at 13.3 AU
and 22.5 AU are close to 3:5 and 2:7 mean resonances with Saturn.
The heliocentric distances of ascending nodes lie mainly in the
range of 6 - 10 AU, therefore the influence of both giant planets,
Jupiter and Saturn, is significant. The position of mean motion
resonances with Saturn do no match perfectly to peaks and gaps of
Lyrid's semimajor axis distribution, but are relatively close.
However, more high quality observational data and following
precise dynamical inspection would be needed for a conclusive
statement.

\section{Conclusions}

The first Modra -- Arboretum double station all-sky video meteor
observation test within the frame of the new Slovak Video Meteor
Network (SVMN) shows reliable results and should provide a good
quality orbits for the future detail studies. The obtained data
are comparable with other known databases (IAU MDC, SonotaCo).

Further improvement of the data quality will be achieved in the
close future by the digital video camera with the higher
resolution. Currently, we are testing the digital CCD camera DMK
41BU02 instead of the analog one. The resolution of the all-sky
video system is then 1280x960 pixels with the frame rate 15 per
second.

The Lyrid meteor stream structure seems to be more complex as it
was considered in previous works. Currently, the number of precise
orbits of Lyrids is not so high for more detail
theoretical studies. The beginning heights of Lyrids do not depend strongly on the absolute brightness. This behavior is similar to Geminids.\\

This work was supported by the Slovak Scientific Grant Agency
VEGA, grant No. 1/0636/09 and Comenius University grant
UK/245/2010. Authors are thankful to the staff of the Astronomical
Observatory in Modra and Arboretum Tes\'{a}rske Mly\v{n}any for
their regular observations and maintenance. Also we are thankful
to Dr. Mikiya Sato for his valuable comments.


\end{document}